\documentclass[floatfix,altaffilletter,superscriptaddress,preprintnumbers,
               tightenlines,showpacs,showkeys,nofootinbib,notitlepage]{revtex4-2}\usepackage[utf8]{inputenc}
\usepackage{amssymb}
\usepackage{amsmath}
\usepackage{graphics}
\usepackage{graphicx}
\usepackage{tikz}
\usepackage[compat=1.0.0]{tikz-feynman}
\usepackage{xcolor}

\usepackage[caption=false]{subfig}

\newcommand{\beq}{\begin{equation}}
\newcommand{\eeq}{\end{equation}}

\newcommand{\dr}{\partial}
\newcommand{\dd}{\mathrm{d}}
\newcommand{\Tr}{\text{Tr}}

\newcommand{\proton}{\mathrm{p}}

\usepackage[normalem]{ulem}

\begin{document}
\title{Chirality distributions inside baryons in ${\rm QCD_2}$}

\author{Adrien Florio}
\affiliation{Center for Nuclear Theory, Department of Physics and Astronomy, Stony Brook University, NY 11794-3800, USA}
\author{David Frenklakh}
\affiliation{Center for Nuclear Theory, Department of Physics and Astronomy, Stony Brook University, NY 11794-3800, USA}
\author{Dmitri E. Kharzeev}
\affiliation{Center for Nuclear Theory, Department of Physics and Astronomy, Stony Brook University, NY 11794-3800, USA}
\affiliation{Physics Department, Brookhaven National Laboratory, Upton, NY 11973-5000, USA}

\begin{abstract}
    The connection between the spin distribution and the topological structure of the baryon is an open and important problem. Here we address it using QCD in $(1+1)$ spacetime dimensions, which is exactly solvable at large number of colors $N$. It is found that the distribution of chirality inside a baryon is drastically different from a chirality distribution inside states with zero baryon number, ``mesons". This difference is shown to arise from the topological structure of the baryon -- at large $N$, all of the baryon's chirality is concentrated near $x=0$, whereas in a meson state it vanishes in the small $x$ limit. Our results illustrate how the constituent features of the baryon reemerge and are tied to the topological features of the bosonized solitonic solution. Possible implications for QCD in $(3+1)$ dimensions and for deep inelastic scattering experiments are discussed.
\end{abstract}

\maketitle

\section{Introduction}

Quantum chromodynamics (QCD) is the established fundamental theory of strong interactions. Nevertheless, the ways in which the properties of hadrons and their interactions emerge from this microscopic theory remain poorly understood. Two of the most interesting open problems are the confinement of quarks and the distribution of the baryon's spin among its constituents. The latter is usually addressed within the parton model that does not fully take into account the nonperturbative structure of the baryon (which only enters through the initial conditions in the Dokshitzer–Gribov–Lipatov–Altarelli–Parisi (DGLAP) parton evolution equations). This lack of connection between the nonperturbative and perturbative sides of the problem may in fact be at the origin of the difficulties in theoretical understanding of the spin distribution inside the proton, or the ``proton spin puzzle" (see \cite{roberts_1990,Manohar:1992tz,Anselmino:1994gn,Filippone:2001ux,Bass:2004xa,Aidala:2012mv} for reviews).

In this paper, we address the connection between the nonperturbative structure of the baryon and the spin (or rather, chirality) distribution inside of it using QCD in $(1+1)$ dimensions, QCD$_2$.  This model is  exactly solvable in the limit of large number of colors $N$ and is known to share many properties with QCD$_4$, including confinement, mass gap generation, and spontaneous breaking of chiral symmetry \cite{tHooft:1974pn,Witten:1983ar,Frishman:1992mr,Gross:1995bp}.

While the notion of spin is absent in one spatial dimension, what is often meant by the ``spin" structure of the proton in high-energy interactions, even in four spacetime dimensions, is really its chirality content, which is perfectly well-defined in two spacetime dimensions. Furthermore, at high energy, dimensional reduction leads to the separation of transverse and longitudinal degrees of freedom, so that ($1+1$)-dimensional dynamics may be more relevant for understanding QCD$_4$ than naively expected. Crucially, large $N$ QCD$_2$ is an exactly solvable  model, valid at all energy scales, which will allow us to address the effect of hadron topological structure on the chirality distribution.

Indeed, the internal topology of the baryon may affect the chirality distribution inside of it. It has long been known  \cite{Skyrme:1962vh, Witten:1979kh} that baryons can be described as topological solitons. The importance of this topological structure for the spin distribution 
 has already been addressed in the context of the Skyrme model \cite{Brodsky:1988ip}. It has been found that the chirality of valence quarks does not contribute to the spin of the baryon, and all of it arises from the topology of the Skyrmion solution. However in $(3+1)$ dimensions, the Skyrme model represents an effective low-energy description of QCD - therefore its validity at short distances, and thus the relevance for the interpretation of the data from deep inelastic scattering (DIS) can be questioned. In contrast, in QCD$_2$ the topological description applies at all energy scales, and will allow us to investigate the effect of baryon topology on its chirality distribution from first principles.

In the 't Hooft limit ($N \to \infty$, $g^2 N$ fixed), QCD$_2$ reduces to an interacting sine-Gordon model \cite{Coleman:1974bu,Faddeev:1977rm,Zamolodchikov:1978xm}. In this bosonic description, the baryon emerges as a topological soliton solution -- a kink.  In this work, we apply ideas from \cite{Dvali:2015jxa} to describe the quantum soliton state in terms of its constituents. 
This decomposition is analogous to the parton model in QCD$_4$, but fully takes account of the topological structure of the soliton. We then use this decomposition to evaluate the chirality distribution inside the soliton. 
\vskip0.3cm


This paper is organized as follows. We motivate our calculation from the proton spin problem point of view in section \ref{sec:motivation}. We then review some  properties of QCD$_2$ and of the sine-Gordon model in sections  \ref{sec:qcd2} and \ref{sec:sg}. Section \ref{sec:results} contains the original results of this work. In section \ref{sec:quantumsol} we extend the results of \cite{Dvali:2015jxa} on quantum solitons to the sine-Gordon model. We then use these results in section \ref{sec:axialcur}  to define and compute the matrix elements of axial current, representing a $(1+1)$ analog of the polarized structure function $g_1(x_B)$ ($x_B$ is the Bjorken $x$). We conclude in section \ref{sec:discussion} with a discussion of these results and their possible implications for the $(3+1)$-dimensional QCD and the spin structure of the baryons. For the sake of self-completeness, in appendix ~\ref{sec:protonspin} there is a pedagogical review of deep inelastic scattering and of the proton ``spin crisis".

\section{The proton spin, topology, and the Sine-Gordon model} \label{sec:prelim}

\subsection{Motivation}
\label{sec:motivation}

Before moving on to $(1+1)$ dimensions, let us briefly recall the role  chirality distributions play in spin distributions. We refer the interested reader to Appendix \ref{sec:protonspin} and references therein for more detailed information about $3+1$ spin physics.

 The matrix elements describing the interaction of hadrons with the electromagnetic field are parametrized in terms of structure functions which need to be measured experimentally or evaluated in lattice simulations. Of its spin dependent parts, the proton polarized structure function $g_1(x_B)$ is of particular interest [see eq. (\ref{eq:SFdefinitions}) for its definition and eq.(\ref{eq:xBdefinition}) for the kinematic definition of the Bjorken $x_B$]. It is experimentally measured in polarized deep inelastic scattering. Moreover, perturbative QCD relates  its first Mellin moment to the axial charges of the nucleon:

\begin{align}
  \int_0^1 \dd x_Bg_1(x_B)  =  \frac{1}{12} a^{3} + \frac{1}{36} a^{8} + \frac{1}{9} a^{0} \ ,
\end{align}

where the axial charges $a^3, a^8, a^0$ are related to the matrix elements of the corresponding QCD currents over the proton state:

\beq
2 M s^\mu a^k \propto \langle \proton,s| j_5^{\mu (k)}| \proton,s\rangle.
\eeq

With $a^3$ and $a^8$ inferred from independent experiments, polarized DIS provides a measurement of the axial charge $a^0$ and polarized quark distributions. In other words, it probes the chirality distribution inside a hadron. It is yet unknown how the different internal structures of baryons and mesons affect their chirality distributions. To make progress toward answering this question, we will study this question in the (1+1) dimensional sine-Gordon model, which is exactly solvable.

\subsection{QCD$_2$ in the 't Hooft limit: The sine-Gordon model}
\label{sec:qcd2}

In order to gain better knowledge about the chirality distribution inside hadrons, we will move on to $1+1$ dimensions and compute this distribution explicitly in QCD$_2$ in the 't Hooft limit. In this section, we give a brief overview of the features of this model that are relevant for our computation. In particular, we recall that its bosonic representation is an interacting  sine-Gordon model.

Consider a non-Abelian gauge field $A_{\mu}\in SU(N)$ minimally coupled to one flavor of fermion $q$ in the fundamental representation of $SU(N)$,
\beq
\mathcal{L}_{QCD_2} = -\frac{1}{4} \Tr F_{\mu\nu}F^{\mu\nu} + i \bar{q} \gamma^\mu(\partial_\mu - i g A_\mu) q - m\bar{q} q \, , \label{eq:QCD2}
\eeq
with $F_{\mu\nu}$ the gluon field strength tensor and $m$ the mass of the fermion. The theory was first considered in the context of the large $N$ expansion \cite{tHooft:1974pn}. It was later realized that this theory, as was already known for the Schwinger model \cite{Schwinger:1962tp}, also admits a dualized bosonic form \cite{Baluni:1980bw, Steinhardt:1980ry}. We refer the interested reader to \cite{Zhitnitsky:1985um, Frishman:1992mr, Abdalla:1995dm} and references therein for more details and reviews on QCD$_2$.

In this work, we will focus on the large $N$, weak coupling 't Hooft limit defined by
\begin{align}
    N\to\infty, \ \ g\to 0, \ \  N\cdot g^2\equiv \lambda = \text{const} \ .
\end{align}
As shown in \cite{Steinhardt:1980ry} and in agreement with the spectrum originally derived in \cite{tHooft:1974pn}, the leading order Lagrangian is nothing but the interacting sine-Gordon model

\beq
\mathcal{L} = \frac{1}{2} (\partial_\mu \phi)^2 - m'^2\cos\left(\frac{\phi}{f}\right) \ , \label{eq:lagSG}
\eeq
with $f = \sqrt{\frac{N}{4\pi}}$. $m'$ is a low-energy parameter which depends on $N, g$ and $m$; it vanishes when $m$ vanishes. Keeping in mind that its precise relation to the original parameters  is  renormalization-scheme dependent, we will use in this work the relation derived in \cite{Steinhardt:1980ry}
\beq
m'^2 = \left[N m \left(\frac{g}{\sqrt{\pi}}\right)^{(N-1)/N} \right]^{2N/(2N-1)} \ ,
\eeq
which corresponds to renormalizing the Hamiltonian by normal ordering with respect to the new mass parameter $m'$. As we shall clarify in the next section, for $1/f\neq 2\sqrt{\pi}$, the sine-Gordon model is equivalent to a model of interacting fermions.

From the bosonization perspective, each color of the original fermion is mapped onto a bosonic phase degree of freedom $\phi^a$. The bosonic field $\phi$ which dominates the physics at large $N$ is the sum of these different phases. Note  that this mode $\phi$ is precisely the one associated with axial rotation of the original fermion.  As remarked in~\cite{Anand:2021qnd}, the Lagrangian \eqref{eq:lagSG} can be thought of as the low-energy chiral effective theory of \eqref{eq:QCD2}. In this context, the  specific value of $f$ (which corresponds to a specific interaction strength) is obtained through 't Hooft anomaly matching; it is fixed by the anomaly coefficient of QCD$_2$.

Let us recall this argument, following \cite{Anand:2021qnd}. Consider for simplicity QCD$_2$ with one flavor of massless quarks. In the low-energy effective theory there is a massless free bosonic field (the ``pion field"). The theory is invariant under a shift symmetry of this field that corresponds to the axial symmetry of the fermionic theory. Explicitly,

\beq
\psi\rightarrow e^{i\alpha \gamma^5}\psi \Longleftrightarrow \phi \rightarrow \phi + 2\alpha f \ .
\eeq
The generator of this symmetry is $Q_5 = \int j_5^0(x) dx$, so the axial current on the bosonic side is defined as

\beq
j_5^\mu = 2 f \partial^\mu \phi \ .
\eeq

Anomaly matching is performed by evaluating a matrix element $\langle j_5^\mu(p) j_5^\nu(-p)\rangle$ both in the IR (free massless boson) and UV (massless quarks) theories. In the IR, the free massless boson theory gives

\beq\label{aIR}
\langle j_5^\mu(p) j_5^\nu(-p) \rangle = 4 f^2 \frac{p^\mu p^\nu}{p^2} \ .
\eeq
In the UV, this matrix element gets a contribution from the anomalous quark loop. The result is proportional to the number of quarks' colors $N$:
\beq\label{aUV}
\langle j_5^\mu(p) j_5^\nu(-p) \rangle = \frac{N}{\pi} \frac{p^\mu p^\nu}{p^2} \ .
\eeq
By matching the anomaly in the IR and UV limits we obtain $f= \sqrt{\frac{N}{4\pi}}$ which is indeed the value used above in the Sine-Gordon Lagrangian (\ref{eq:lagSG}). It also matches the value obtained from bosonization \cite{Steinhardt:1980ry}, and agrees with the scaling of the pion decay constant in QCD$_4$.

\subsection{Properties of the sine-Gordon model}
\label{sec:sg}

The sine-Gordon model, given by the Largangian in eq. (\ref{eq:lagSG}), is very well-studied due to its classical and quantum integrability; we refer the interested reader  to \cite{Coleman:1974bu, Dashen:1975hd, Faddeev:1977rm} and references therein. One of the intriguing properties of the (bosonic) sine-Gordon model is that it is the bosonic dual of the  massive Thirring model, given by the Lagrangian \cite{Coleman:1974bu, Faddeev:1977rm}

\beq
\mathcal{L} = \bar{\psi}i\gamma^\mu\partial_\mu\psi - m_T \bar{\psi}\psi - \frac{1}{2}g_T (\bar{\psi}\gamma^\mu\psi)^2 \ .
\eeq
The coupling constant $g_T$ of the Thirring model is related to the period of the sine-Gordon boson, which is defined by $f$ in eq. (\ref{eq:lagSG}):

\beq
1+\frac{g_T}{\pi} = 4\pi f^2 \ .
\eeq

In particular, $f = \dfrac{1}{2\sqrt{\pi}}$ corresponds to $g_T=0$, namely a free fermion. In our case of large $N$, $f = \sqrt{\dfrac{N}{4\pi}}$ and it corresponds to a finite (large) coupling constant in the Thirring model.

Another relevant property of  the sine-Gordon model is the appearance of topologically nontrivial classical solutions - kinks and antikinks. The solution to classical equations of motion of the Lagrangian (\ref{eq:lagSG}) which describes a stationary kink located at the point $x=0$ is

\beq \label{eq_kink}
\phi_c(x) = \sqrt{\frac{4 N}{\pi}} \arctan e^{\sqrt{\frac{4\pi}{N}}m' x} \ .
\eeq
Classical mass of the kink is computed by calculating the energy of this solution. The Hamiltonian corresponding to our Lagrangian (\ref{eq:lagSG}) is

\beq \label{eq:class_Ham}
H = \int dx \left[\frac{1}{2} (\partial_t \phi)^2 + \frac{1}{2} (\partial_x \phi)^2 + m'^2 \left(1 - \cos\left[\sqrt{\frac{4\pi}{N}}\phi\right]\right)\right] \ .
\eeq
Note that for convenience we added a constant term compared to the Lagrangian (\ref{eq:lagSG}), which of course does not affect the physics but makes the vacuum energy set to zero, so the energy we are computing is just the mass of the kink. The classical kink profile satisfies a Bogomol'nyi–Prasad–Sommerfield (BPS) condition, which relates its gradient energy to its potential energy:
\beq \label{eq:BPS}
\frac{1}{2}(\partial_x\phi_c)^2 = m'^2 \left(1 - \cos\left[\sqrt{\frac{4\pi}{N}}\phi_c\right]\right) \ .
\eeq

The BPS condition arises from the fact that the stationary kink solution minimizes energy in the given topological sector. By plugging the solution (\ref{eq_kink}) into the Hamiltonian (\ref{eq:class_Ham}) we obtain the classical mass of the kink:
\beq \label{eq_m_kink}
M_{kink} = 4\sqrt{\frac{N}{\pi}} m'
\eeq

It is also instructive to compute the vector charge of the kink. Using bosonization, the vector current reads
\beq
j^\mu = \sqrt{\frac{N}{\pi}}\epsilon^{\mu\nu} \partial_\nu \phi \ ,
\eeq
and leads to the total vector charge
\beq
Q = \sqrt{\frac{N}{\pi}}\left[\phi(x\rightarrow\infty) - \phi(x\rightarrow -\infty)\right] \ .
\eeq
We see that the fermionic vector charge is mapped onto a topological charge in the bosonic language. Moreover, for the kink solution (\ref{eq_kink}) this gives charge $N$; this shows that the kink consists of $N$ quarks and represents a bosonic description of a baryon.

Note also that in 1+1 dimensions the axial and vector currents are related to each other

\beq\label{vec_ax}
j^\mu_5 = \epsilon^{\mu\nu}j_\nu \ .
\eeq
In other words, the vector charge density is the axial current density. The vector charge in our model is the topological charge, which thus defines the axial current. We have seen that in (3+1) QCD,  the axial current matrix element over a nucleon state is related to the nucleon's spin. We see that in our model the role of spin is played by chirality which arises from the topology of the field configuration. 

We have so far only discussed classical profiles. A legitimate question to ask is how quantum corrections affect these solutions. A remarkable property of the sine-Gordon model is that it is ``one-loop exact" \cite{Dashen:1975hd, Faddeev:1977rm}. In the case of the kink one-loop correction only amounts to shifting its mass \eqref{eq_m_kink}, see \cite{Dashen:1975hd}
\beq
M_{kink}^{1-loop} = \left(1 - \frac{1}{2N}\right) M_{kink}\ .
\eeq
In the limit $N\to \infty$ we are considering, it is thus enough to consider classical kinks.

\section{Chirality of topological solitons in the Sine-Gordon model: exact results} \label{sec:results}

\subsection{Solitons as quantum states}
\label{sec:quantumsol}

In order to consider the matrix element of axial current in a baryon state, we first need to define what a baryon state is. As explained in the previous section, on the bosonic side, the baryons are mapped onto classical topological solitons. In this section, we will quantize these solitons. Namely, we will construct quantum states whose expectation values are the classical soliton solutions. To achieve this, we  follow the method of \cite{Dvali:2015jxa}, where solitonic states and encoding of topology in these states were studied in a scalar $\phi^4$ theory.

Explicitly, we seek a state $\left | kink \right>$ such that
\beq \label{eq_state}
\langle kink| \hat{\phi}(x) | kink \rangle = \phi_c (x)
\eeq
where $\hat\phi$ is the field operator and $\phi_{c}$ is the kink solution defined in \eqref{eq_kink}. Consider the Fourier decomposition of $\phi_c(x)$
\beq \label{eq_Fourier_coord}
\phi_c(x) = \int \frac{dk}{2\pi} \frac{1}{\sqrt{2\omega(k)}} (\alpha_k e^{ikx} + \alpha^*_k e^{-ikx}) \ ,
\eeq
where $\alpha_k$ are Fourier coefficients and $\omega(k)$ is a yet unspecified function of $k$ that will play the role of a dispersion relation. This formula is introduced to mimic a free field expansion in Fourier space. Explicitly, taking the Fourier transform of \eqref{eq_kink}, we have \footnote{In the following result we have neglected a contact term proportional to $\delta(k)$ which arises due to the fact that our kink is not centered around $\phi=0$. A redefinition of $\phi$ by a constant shift would get rid of the contact term explicitly, but we prefer to use a more conventional sine-Gordon Lagrangian. We just note that none of our results will depend on this contact term so it is dropped from here on. }

\beq \label{eq:alphak}
\alpha_k = - \sqrt{2\pi N\omega(k)}\frac{i}{2k} \frac{1}{\cosh\left(\sqrt{\frac{N}{4\pi}}\frac{\pi k}{2 m'}\right)} \ .
\eeq
This suggests the following decomposition of our field operator:
\beq \label{eq:phi_decomp}
\hat{\phi}(x) = \int \frac{dk}{2\pi}\frac{1}{\sqrt{2\omega(k)}} ( a^{sol}_k e^{ikx} + a^{sol\dagger}_k e^{-ikx}), \
\eeq
with $a^{sol}_k$ an operator satisfying the property
\beq
\label{eq:def_prop_a}
\langle kink| a^{sol}_k| kink \rangle = \alpha_k \ .
\eeq
As this expansion suggests, the operators $a^{sol}_k, a^{sol\dagger}_k$ are a set of annihilation and creation operators. Indeed, we can define the  momentum operator conjugate to $\hat{\phi}$ by
\beq \label{eq:pi_decomp}
\hat{\pi}(x) = \int \frac{dk}{2\pi}(-i)\sqrt{\frac{\omega(k)}{2}} (a^{sol}_k e^{i k x} - a^{sol\dagger}_k e^{-i k x})
\eeq
provided that
\beq
 [a_k^{sol}, a_{k'}^{sol\dagger}] = 2\pi\delta(k-k')
\eeq
so that the canonical commutation relation
\beq
[\hat\phi(x), \hat\pi(x')] = i \delta(x-x')
\eeq
is satisfied.

It is important to emphasize that the operators $a^{sol}_k, a^{sol\dagger}_k$ are \textit{not} the creation and annihilation operators of the noninteracting theory. In particular, they do \textit{not} diagonalize the Hamiltonian on their associated vacuum. 

From equation \eqref{eq:def_prop_a}, we see that the kink state we are looking for is nothing but a combination of  coherent states of these creation and annihilation operators. Indeed, the state $|\alpha_k\rangle$
\begin{align}
|\alpha_k\rangle &= e^{-\frac{1}{2}|\alpha_k|^2} e^{\alpha_k a_k^{sol\dagger}}|0_k^{sol}\rangle\\
&=e^{-\frac{1}{2}|\alpha_k|^2}\sum_{n_k=0}^\infty \frac{\alpha_k^{n_k}}{\sqrt{n_k!}}|n_k\rangle
\end{align}
is a coherent state associated with $a_k^{sol}$, $a_k^{sol} |\alpha_k\rangle = \alpha_k |\alpha_k\rangle$. The state $|0_k^{sol}\rangle$ is the vacuum state annihilated by $a_k^{sol}$. Correspondingly, $|n_k^{sol}\rangle=\frac{(a_k^{sol\dagger})^{n_k}}{\sqrt{n_k!}}|0_k^{sol}\rangle$ is the state with occupation number $n_k^{sol}$; the state of ``$n$ solitonic constituents" with momentum $k$. We then easily see that the state
\begin{align}
  |kink\rangle = \bigotimes_k |\alpha_k\rangle
\end{align}
satisfies the sought-after property \eqref{eq_state}.

Before moving on, it will be instructive to study some properties of this kink state. First let us show that, upon imposing the correct dispersion relation $\omega(k)$, the energy of the kink is recovered by summing the energies of its constituents.
The classical Hamiltonian of the sine-Gordon model is given by eq. (\ref{eq:class_Ham}). It consists of kinetic energy, gradient term and potential term. The potential term is nonquadratic and in general prevents from diagonalizing the Hamiltonian operator by introducing creation/annihilation operators. However, in the kink state, the BPS condition (\ref{eq:BPS}) is  satisfied, and the potential term can be traded for another gradient term. We assume that the Hamiltonian is normal ordered with respect to the solitonic constituent creation-annihilation operators, so that the vacuum energy is zero. As a result, on the kink state, the BPS condition vastly simplifies the Hamiltonian:
\beq
\langle\hat{H}\rangle_{kink} = \langle kink |\hat{H}|kink\rangle = \langle kink|\int dx \left[ \frac{1}{2} (\hat{\pi})^2 + (\partial_x \hat{\phi})^2\right] |kink\rangle \ .
\eeq
Note that this expression resembles a free Hamiltonian except for the coefficient of the gradient term. This difference can be traced back to the use of  the BPS condition to convert the cosine term into the gradient term.
 Now using the expressions (\ref{eq:phi_decomp}) and (\ref{eq:pi_decomp}) we obtain an expression for the expectation value of the Hamiltonian over the kink state in terms of creation and annihilation operators:
\beq
\langle\hat{H}\rangle_{kink} = \left \langle \int \frac{dk}{2\pi} \left\{(a^{sol}_k a^{sol}_{-k} + a^{sol\dagger}_k a^{sol\dagger}_{-k}) \left[\frac{k^2}{2 \omega(k)} - \frac{\omega(k)}{4}\right] + 2 a^{sol\dagger}_k a^{sol}_k \left[\frac{k^2}{2 \omega(k)} + \frac{\omega(k)}{4}\right] \right\} \right \rangle_{kink} \label{eq:kink_H} \ .
\eeq
We are now in a position to choose $\omega(k)$ such that the unwanted contributions cancel from the Hamiltonian.
It leads to \footnote{Note that this result differs by a factor of $\sqrt{2}$ from the one given in \cite{Dvali:2015jxa}.}
\beq
\omega (k) = \sqrt{2}|k|.
\eeq
This expression for the dispersion relation shows again that the excitations created by $a_{k}^{sol\dagger}$ are not free particle excitations. They are ``corpuscular constituents" which make up coherently our quantum soliton state.

With this  dispersion relation at hand, the Hamiltonian (\ref{eq:kink_H}) can be rewritten in terms of occupation number operator $\hat{N}^{sol}_k = a^{sol\dagger}_k a^{sol}_k$. The kink mass (\ref{eq_m_kink}) can therefore be calculated by summing the energies over all the constituents with the corresponding average occupation numbers $N_k = |\alpha_k|^2$
\beq
M_{kink} = \int \frac{dk}{2\pi} \omega(k) N_k = \int \frac{dk}{2\pi} \omega(k) \cdot |\alpha_k|^2 \ .
\eeq
Plugging in \eqref{eq:alphak}, we indeed recover the expression for the kink mass (\ref{eq_m_kink}).

As first realized in \cite{Dvali:2015jxa}, an interesting feature of expressing a kink as a quantum state is the possibility of understanding the appearance of topology at the corpuscular level. This is done by splitting  the Fourier coefficients \eqref{eq:alphak} into  ``topology" and ``energy" parts (remember that a contact term in $\alpha_k$ is dropped, see the discussion under a footnote before eq. \eqref{eq:alphak}) 
 
\beq
\alpha_ k  = t_k c_k ~~, ~~ t_k = \frac{i \sqrt{\omega(k)}}{k} ~~,~~ c_k = -\sqrt{\frac{\pi N}{2}}\frac{1}{\cosh{\sqrt{\frac{N}{4\pi}}\frac{\pi k}{2m'}}} \ .
\eeq
Equivalently, the field configuration in position space is represented as a convolution of ``topology" and ``energy" 

\beq \label{eq_convolution}
\phi_c(x) = \sqrt{\frac{N}{\pi}} (\text{sign} \ast  \text{sech}) \left(\sqrt{\frac{N}{4\pi}}m' x\right)  + \frac{\pi}{2}  \ ,
\eeq
where the $\text{sign}$ function is defined as $\pm\frac{1}{2}$ depending on the sign of the argument. It is easy to verify that the presence of the sign function is crucial to get a nonzero topological charge. In Fourier space, it translates into a singularity at $k=0$. In this language, topology can be understood as a condensation of constituents in the zero mode. It reflects the global aspects of topological configuration and will play a crucial role in the rest of this work. The additional constant $\frac{\pi}{2}$ is at the origin of the contact term and is irrelevant for our computation.

We also note that, surprisingly, the energy part of our sine-Gordon soliton is made of a nontopological soliton of the inverted $\lambda\phi^4$ of \cite{Dvali:2015jxa}.  In this theory, the soliton profile takes the form

\beq \label{eq_nontop}
\phi_{n-t} (x) = \frac{m}{\sqrt{\lambda}} \text{sech}(mx) \ ,
\eeq
with $m$ the bare mass of the theory. This suggests that such a decomposition between energy and topology is indeed meaningful and may help shed some new light on the structure of topological states.

\subsection{Axial current on the light cone}
\label{sec:axialcur}

To make contact with the traditional parton model, let us now formulate our model in the infinite momentum frame. This will allow us to neglect the baryon mass, in accord with the parton model.  To go to the infinite momentum frame, we will make a boost with $\beta\rightarrow 1$. 
 The field profile becomes

\beq \label{eq:boosted_kink}
\phi_b (x,t) = \sqrt{\frac{N}{4\pi}} 4 \arctan e^{\sqrt{\frac{4\pi}{N}}m' \gamma (x + \beta t)} \xrightarrow[\beta\rightarrow\ 1]{} \sqrt{\frac{N}{4\pi}} 4 \arctan e^{\sqrt{\frac{4\pi}{N}}m' \gamma \sqrt{2} x^+}
\eeq
where we have defined light-cone coordinates $x^{\pm} = \dfrac{1}{\sqrt{2}} (t\pm x)$, and the $\gamma$ parameter is as usual $\gamma = \dfrac{1}{\sqrt{1-\beta^2}}$. For a proton on the light-cone we consider a decomposition of the field operator into light-cone constituents:

\beq
\hat{\phi}(x^+) = \int \frac{dk_+}{2\pi}\frac{1}{\sqrt{2 \sqrt{2} k_+}} ( a_{k_+} e^{i k _+ x^+} + a_{k_+}^\dagger e^{-i k_+ x^+})
\eeq

Following the same steps outlined above for a stationary kink, we could introduce a boosted kink state as a coherent state of such constituents with the corresponding coefficients $\alpha_{k_+}$. Similar to eq. (\ref{eq_Fourier_coord}) $\alpha_{k_+}$ could be calculated as the light-cone Fourier coefficients of the classical field configuration. As a result, we obtain

\beq
\alpha_{k_+} = \sqrt{2 \sqrt{2} k_+ N  \pi}\frac{i}{2k_+} \frac{1}{\cosh\left(\sqrt{\frac{N}{4\pi}}\frac{\pi k_+}{2 \sqrt{2} \gamma m'}\right)} \ .
\eeq

It is instructive to rewrite this expression using as a variable Bjorken $x_B$. Note that the + component of proton's momentum is $p_+ = \sqrt{2} \gamma M_{p}$ where $M_{p}$ is the proton mass given by the expression (\ref{eq_m_kink}). Since solitonic constituents play the role of partons, $x_B = \dfrac{k_+}{p_+}$. Therefore, the light-cone Fourier coefficients can be rewritten in terms of $x_B$ as

\beq
\alpha_{k_+} = \sqrt{2 \sqrt{2}N x_B p_+ \pi}\frac{i}{2 x_B p_+} \frac{1}{\cosh\left[N x_B\right]} \ .
\eeq
We see that the constituent occupation number
\beq
N_{k_+} = |\alpha_{k_+}|^2 = \frac{\pi N}{\sqrt{2} x_B p_+} \frac{1}{\cosh\left[ N x_B\right]^2}
\eeq
has a logarithmic divergence at small $x_B$ due to the kink's topological structure.

\begin{figure}
     \includegraphics[scale=0.35]{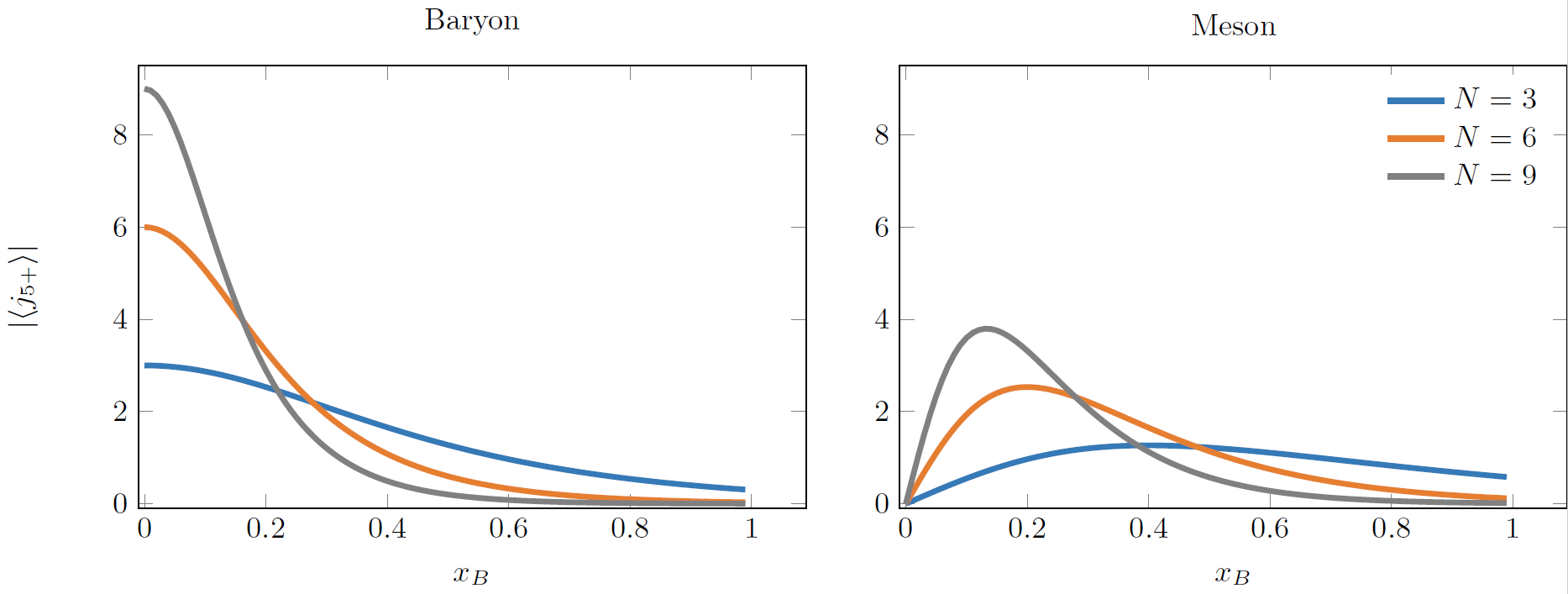}
    \caption{\textbf{Left panel:} The matrix element of the axial current inside a baryon as a function of Bjorken $x$ for different numbers of colors $N$. For large $N$, the $x_B\sim 1$ region is suppressed while the $x_B\rightarrow 0$ region is enhanced. 
    \textbf{Right panel:} Same as left panel, but for a meson.  The $x_B\rightarrow 0$ region contribution vanishes, reflecting the lack of baryon topology. }
    \label{plot:ax_current}
\end{figure}

We are now ready to calculate the matrix element of the axial current in the state of an infinitely fast proton. Remembering from the bosonization dictionary that $j^\mu_5 = \sqrt{\dfrac{N}{\pi}} \partial^\mu \phi$ and using properties of coherent states in position space we get

\beq
\langle kink | j_{5+} (x^+) | kink \rangle =  \sqrt{\frac{N}{\pi}} \frac{2\sqrt{2}m'\gamma}{\cosh\left(\sqrt{\frac{8\pi}{N}}\gamma m' x^+\right)} = \frac{p_+}{2\cosh\left(\frac{\pi p_+}{2N} x^+\right)} \ .
\eeq
Going to the momentum space and expressing the results in terms of Bjorken $x_B$ we get from the above

\beq \label{eq:ax_current}
 \langle kink |j_{5+}(x_B) | kink \rangle = N \frac{1}{\cosh\left[N x_B\right]} \ .
\eeq
This expression shows that the matrix element of the axial current is dominated by the low $x_B$ region, $x_B \sim \frac{1}{N}$. We illustrate this in  Fig. (\ref{plot:ax_current}). For large enough $N$ there is a strong suppression of the region $x_B\sim 1$ while at the same time the small $x_B$ region is enhanced.

This behavior at low $x_B$ can be understood as arising from the nontrivial topology of the baryon. We can show this explicitly by considering the decomposition of our soliton state \eqref{eq_convolution} into an energy component and a topological component. Rewriting it in the infinite momentum frame, we have

\beq
\phi_b (x^+) = \sqrt{\frac{N}{\pi}}(\text{sign} \ast \text{sech}) \left(\frac{\pi p_+ x^+}{2N}\right) \ .
\eeq
To mimic the meson wave function, we remove the baryon topology by 
replacing the sign function by a $\delta$-function and consider only the energy part
\beq
\phi_b^{n-t} (x^+) = \sqrt{\frac{N}{\pi}} \frac{1}{\cosh\left(\frac{\pi p_+ x^+}{2N}\right)} \ .
\eeq
Of course, this solution is not a classical solution of the equation of motion; it is meant to provide a caricature of the meson wave function that does not possess the baryon topology.

Going through the steps outlined above, we find the axial current in the coherent state corresponding to this nontopological ``mesonic" soliton:
\beq
\langle nontop.|j_{5+}(x_B)|nontop.\rangle (x_B) = \frac{2N^2 x_B i}{\pi \cosh(N x_B)} .
\eeq
We see that at small $x_B$ this expression vanishes, in stark contrast to the case of the baryon.  Comparing the left and right panels of  Fig. \ref{plot:ax_current}, we clearly see that the $x_B\rightarrow 0$ region contains only the topological contribution. The energy contribution vanishes at small $x_B$.

\section{Discussion}
\label{sec:discussion}
In the traditional parton model, parton distributions are sensitive to the nonperturbative hadron structure only through the initial conditions for the evolution equations. The perturbative QCD evolution is then completely decoupled from the hadron structure, and is identical for baryons and mesons.  There is, however, an important question about the role of nonperturbative effects, and in particular the effect of baryon topology on parton distributions. Addressing this question is very hard in real  $(3+1)$-dimensional QCD, but we have solved the problem of parton distributions inside baryons exactly using ${\rm QCD}_2$ in the 't Hooft limit. 
\vskip0.3cm

In this limit, ${\rm QCD}_2$ reduces to the interacting sine-Gordon model. As we saw in the previous section, its topological solitons are dual to baryons. Moreover, they can be realized as quantum coherent states. We have evaluated the matrix element of the axial current on such states on the light cone. The soliton wave function can be exactly decomposed in a convolution of a topological and a nontopological contribution. We have shown that the main contribution (at leading order in $1/N$, where $N$ is the number of colors) to the ``baryon" expectation value of $j_{5+}(x_B)$ comes from the region $x_B<\frac{1}{N}$.  We have also explicitly demonstrated that this enhancement at $x_B=0$ originates from the topology of the soliton. 

\vskip0.3cm

These results provide new insights on the baryon structure at large $N$. Even though they have been obtained through bosonization, they are consistent with the  picture of $N$ quarks moving in a mean field created by weak $\sim 1/N$ binary potentials, with $N^2$ quark pairs contributing to the average Hartree potential \cite{Witten:1979kh}. We find that this picture describes not only the mass of baryons at large $N$, but also their chirality distributions -- namely, each quark carries $\sim 1/N$ fraction of the baryon's chirality. Our results illustrate how the constituent quark picture of  the baryon reemerges and is tied to the topological features of the bosonized solitonic solution.


\vskip0.3cm
One may wonder whether our results obtained in QCD$_2$ are relevant in the real $(3+1)$-dimensional world. The potential relevance of these results for spin physics follows from the operator product expansion (OPE) \eqref{eq:OPEg1} that relates the hadronic tensor in polarized DIS to the expectation value of the axial current.  However, the relation (\ref{vec_ax}) between the axial and vector currents that locks the axial current to the vector charge, and many of the derivations in Section \ref{sec:results} linking the matrix element of the axial current to the topology of the baryon, are specific to $(1+1)$-dimensional theory. Nevertheless, we believe that our results may be relevant for the real QCD, for the following reasons. First, in (3+1) dimensions,
there is also a linking of the UV and IR regions, through the chiral anomaly, with a corresponding anomaly matching; this is reminiscent of the anomaly matching in eqs.  (\ref{aIR}) and (\ref{aUV}). Because of this, the polarized structure functions, even at large momentum transfer, are sensitive to the IR description of the hadron, and thus to its topology. Second, at high energies the transverse and longitudinal degrees of freedom factorize, so our $(1+1)$-dimensional treatment may be more relevant for high-energy interactions in the $(3+1)$-dimensional world than naively expected. 
\vskip0.3cm
One of the big differences between QCD$_2$ and QCD$_4$ is the absence of dynamical gluons in QCD$_2$. In the real $(3+1)$-dimensional QCD, the dynamical gluons are known to contribute to the polarized structure functions via the chiral anomaly \cite{Adler:1969gk,Bell:1969ts} that links the short and large distances \cite{Dolgov:1971ri}. As a result, the matrix element of the axial current measured in DIS becomes sensitive to the large distance, nonperturbative hadron structure \cite{Jaffe:1989jz,Hatsuda:1988jv,Efremov:1989sn,Shore:1990zu,Tarasov:2020cwl,Tarasov:2021yll}. How would the topological structure of the baryon affect the gluon distributions in QCD$_4$? One possible mechanism is offered by the presence of ``baryon junctions" in the wave functions of the baryons that are required by the gauge invariance \cite{Rossi:1977cy}. These junctions have been found to provide an efficient non-perturbative mechanism for baryon stopping \cite{Kharzeev:1996sq}, but their implications for the polarized gluon distributions were never studied.

\vskip0.3cm
What are the phenomenological consequences of our results? Our main prediction is the dramatic difference between the spin distributions inside baryons and mesons. Since the singlet axial current contributes to the polarized structure function  $g_1(x_B)$, we expect that this structure function at small $x_B$ is enhanced  for baryons and suppressed for mesons. This prediction can be tested in experiments, since meson structure functions can be accessed through the diffractive DIS with a baryon in the target fragmentation region separated by a rapidity gap from the inelastic final state. The polarized structure functions of baryons and mesons can also be measured in lattice QCD (see e.g. \cite{gockeler1996polarized,best1997pi,green2014nucleon,chen2016nucleon,lin2018parton}). 
\vskip0.3cm

Another interesting direction is suggested by the Gribov-Lipatov reciprocity relation \cite{gribov1971deep} between the fragmentation functions and parton distributions. For quark fragmentation to protons, the corresponding fragmentation function $D_{p/q}(z)$ is softer than for mesons \cite{particle2020review}. Through the reciprocity relation, this implies a softer quark distribution in Bjorken $x$ inside the proton, in accord with our findings. It would be interesting to extend the present study to spin-dependent quark fragmentation functions $\Delta D_{p/q}(z)$ into protons and vector mesons in semi-inclusive DIS.
\vskip0.3cm

Recently, the relation between the parton model and the hadron structure has been reexamined from the viewpoint of quantum information. Namely, the parton distributions were argued to arise from the reduced density matrix obtained from the pure density matrix of the hadron by tracing over the degrees of freedom that are unobservable in DIS. This approach predicts the emergence of the maximally entangled state at small $x$ \cite{Kharzeev:2017qzs,Kharzeev:2021nzh,Zhang:2021hra} that can be viewed as an alternate picture of saturation \cite{Dvali:2021ooc,Liu:2022ohy,Liu:2022hto} and is supported by the recent analyses \cite{Kharzeev:2021yyf,Hentschinski:2021aux} of the DIS data from HERA \cite{H1:2020zpd}. From this viewpoint, the present work can be viewed as a first step in the extension of this approach to polarized structure functions, where the chiral anomaly plays a crucial role. It would be interesting to extend our study by explicitly evaluating the entanglement entropy and exploring its relation to the polarized structure functions. 
Another future direction would be to try and extend our approach to the $(3+1)$-dimensional QCD on the light cone; one way to proceed would be to use the $N_f = 1$ baryon model proposed in \cite{Komargodski:2018odf}. Within QCD$_2$, it would be interesting to study the effect of $1/N$ corrections, and consider the theory with several quark flavors. 
\vskip0.3cm

\section*{Acknowledgments} 
The authors are grateful to V. Korepin for his patient explanations of the sine-Gordon model and for sharing his insights with us. 
This work was supported by the U.S. Department of Energy,
Office of Science grants No.
DE-FG88ER41450 and No. DE-SC0012704.

\appendix

\section{Proton spin}
\label{sec:protonspin}

This appendix contains a brief overview of the proton spin problem and its connection to gauge field topology, and the ways in which the proton spin distributions are measured in polarized deep inelastic scattering. Our notations follow  \cite{Manohar:1992tz}.

The ``proton spin" measured in DIS experiments is defined through the matrix element of the axial vector current:
\begin{equation}\label{spin_def}
 2M s^\mu = \langle \proton,s|\bar\Pi\gamma^\mu\gamma^5\Pi|\proton,s\rangle \ ,
\end{equation}
where $|p,s\rangle$ is the proton state with spin $s$ and $\Pi$ is the fermionic proton operator; see also \cite{Shore:2007yn} for a more detailed discussion. In the experiment with polarized particles, a nonvanishing matrix element of the axial current manifests itself  through the cross section dependence on polarization. The process most studied in this context is polarized deep inelastic scattering ${\vec l}+{\vec \proton} \to {\vec l}+X$, where ${\vec l}$ is a polarized lepton and $X$ represents the sum over all possible channels. The corresponding cross section depends both on the polarization of the proton and of the initial lepton.

\begin{figure}
    \centering
    \subfloat[Deep inelastic scattering.]{\label{fig:DIS}\includegraphics[scale=0.95]{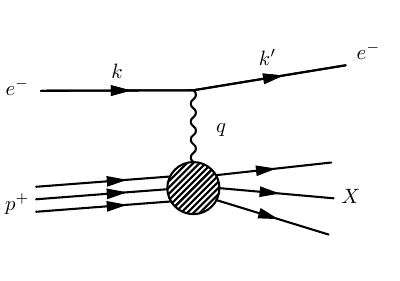}}
    \subfloat[{The photon-parton interaction vertex. $\xi\in  [0,1]$ is the momentum fraction of the parton.}]{\label{fig:DISParton}
  \includegraphics{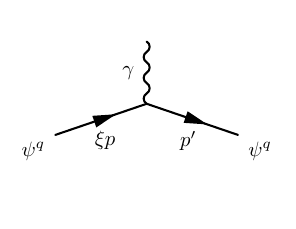}%
}
	\caption{~}
\end{figure}

The leading contribution to DIS is shown in Fig. \ref{fig:DIS}.  Given an incoming lepton with four-momentum $k$ and energy $E$ in the target rest frame, scattering off a hadron of mass $M$ with a momentum transfer $q=k-k'$ and energy loss $\nu=E-E'$, with $k'$ and $E'$ the four-momentum and energy of the outgoing lepton, one can define dimensionless Bjorken $x$ and inelasticity $y$ as
\begin{align}
 \label{eq:xBdefinition} x_B &= \frac{Q^2}{2M \nu} = \frac{Q^2}{2p\cdot q} \ , \\
  y &= \frac{\nu}{E} \ ,
\end{align}
with $Q^2=-q^2$. 
The corresponding differential cross section can be  expressed in terms of the leptonic and hadronic tensors, $l_{\mu\nu}$ and  $W_{\mu\nu}$ (to be defined below), as
\begin{equation}
  \frac{\dd^2 \sigma}{\dd x_B \dd y \dd \phi} = \frac{e^4}{16 \pi^2 Q^4} l_{\mu\nu}  W^{\mu\nu}_{\lambda \lambda} \ ,
\end{equation}
where $\phi$ is the azimuthal angle of the scattered lepton.

The leptonic and hadronic tensors are defined in terms of the correlation functions of the electromagnetic current $j^\mu$ by computing the diagram \ref{fig:DIS} ( see \cite{Manohar:1992tz} for a detailed derivation):
\begin{align}
  l^{\mu\nu} &= \sum_{s_f} \langle k', s_f|j^\nu_l(0)|k,s_l\rangle\langle k,s_l|j_l^\mu(0)|k', s_f\rangle \\
  W^{\mu\nu}_{\lambda\lambda'} &= \frac{1}{4 \pi} \int \dd^4 x e^{iqx} \langle \mathrm{p},\lambda'|[j^\mu(x),j^\nu(0)]|\mathrm{p},\lambda\rangle \ . \label{eq:had_tens_def}
\end{align}

This decomposition fully splits the electromagnetic sector contribution from the QCD contribution. The tensor $l^{\mu\nu}$ can be directly computed in perturbation theory. On the other hand, the hadronic tensor $W^{\mu\nu}_{\lambda\lambda'}$ is  difficult  to evaluate because of the nonperturbative nature of the strong interaction. To study it further, it is useful to decompose the hadronic tensor into an irreducible set of Lorentz structures, the so-called structure functions. In order to proceed, we first need to specify the polarization of the target proton. The most general polarization state for a spin $1/2$ target in its rest frame can be described by the following density matrix
\begin{align}
  \rho = \sum_{\lambda,\lambda'=-1/2}^{1/2} A_{\lambda\lambda'} |\lambda\rangle\langle\lambda'| \ .
\end{align}
In this case, it is convenient to write the matrix $A$ in terms of Pauli operators
\begin{align}
  A =\frac12 \left( \mathbf{1} + \frac{1}{M}\vec s_H\cdot\vec\sigma\right) \ ,
\end{align}
where we used the fact that $\mathrm{Tr}(A) = 1$ and the factor of $M$ is introduced for dimensional reasons. 

The relation between the four-vector $s_\mu$ in (\ref{spin_def})  and $\vec s_H$ is established in the rest frame of the target through $s_\mu  = (0,\frac{\vec s_H}{M })$; the polarization of the target is fully determined by this quantity. Assuming the polarization of the hadron is fixed, the hadronic tensor of interest becomes
\begin{align}
  W^{\mu\nu} = \sum_{\lambda \lambda'} A_{\lambda \lambda'} W^{\mu\nu}_{\lambda \lambda'} \ .
\end{align}
 We are now ready to define the structure functions. Taking into account current conservation $q_\mu W^{\mu\nu}  = q_\nu W^{\mu\nu}  = 0$ together with  the transformation properties of $W^{\mu\nu}$ under time reversal and parity transformations (which are derived from \eqref{eq:had_tens_def}, see for instance \cite{Manohar:1992tz}), the most general Lorentz covariant hadronic tensor for deep inelastic scattering on a polarized spin $1/2$ target is
 \begin{align} \label{eq:SFdefinitions}
   W^{\mu\nu} & = \left(-\eta_{\mu\nu}+\frac{q_\mu q_\nu}{q^2}\right) F_1 +\frac{1}{p\cdot q}\left(p_\mu - \frac{p\cdot q q_\mu}{q^2}\right)\left(p_\nu - \frac{p\cdot q q_\nu}{q^2}\right) F_2 \nonumber\\
   & + \frac{i}{p\cdot q} \epsilon_{\mu\nu\rho\sigma} q^\rho s^\sigma g_1 + \frac{i}{(p\cdot q)^2}\epsilon_{\mu\nu\rho\sigma} q^\rho\left(p\cdot q s^\sigma - s\cdot q p^\sigma\right) g_2 \ .
 \end{align}
The functions $F_1, F_2, g_1, g_2$ are  called  the structure functions. They depend on the relativistic invariants $p^2 = M^2, p\cdot q, Q^2= -q^2$. It is conventional to express this dependence in terms of $Q^2$ and the dimensionless variable $x_B$. The functions $g_1$ and $g_2$ are sensitive to the polarization of the target and are often referred to as polarized structure functions.

When considering high energies and large momentum transfers, a particularly useful tool is the operator product expansion (OPE), which allows one  to  represent two point correlators, such as $W^{\mu\nu}$, as an expansion in local operators\footnote{The OPE is a central tool of perturbative QCD; we refer the interested reader to \cite{Sterman:1995fz} and references therein.}. The power counting for the expansion of $W^{\mu\nu}$ is done in terms of the ``twist" of a given operator
\begin{align}
  t = \mathrm{dimension - spin}
\end{align}
and the leading order are twist two operators. In particular, the OPE allows us to understand the different structure functions in terms of contributions from local operators. In particular, at leading order, the part of the operator expansion of $W_{\mu\nu}$ contributing to $g_1$ (i.e. with the appropriate symmetries) is \cite{Shore:2007yn}
\begin{equation}
  W^{\mu\nu} \underset{\substack{Q^2\to \infty }}{\sim} 2 \epsilon^{\mu\nu\lambda\rho}\frac{p_\lambda}{Q^2}\left[C^{NS}\left(j_{\rho5}^3+\frac{1}{\sqrt{3}}j_{\rho5}^8\right)+\frac{2\sqrt{2}}{\sqrt{3}}C^{S}j^0_{\rho5}\right] \ , \label{eq:OPEg1}
\end{equation}
where $j_{\mu5}^a= \bar{q}\gamma_\mu \gamma_5 T^a q$ is the QCD axial current with $T^a$ ($a = 1,...,8$) being the generators of $SU(3)$, normalized so that $\Tr\left(T^aT^b\right)=\frac{1}{2}\delta^{ab}$; $T^0 = 1 / \sqrt{6}$. $C^{NS}$ and $C^S$ are Wilson coefficients which can be computed in perturbation theory. This relation provides a direct link between the structure function $g_1$ and the miscroscopic QCD axial current. In other words, measurements of $g_1$ can be interpreted as probing the axial charge content of the proton. This is also why measurements of $g_1$ are often referred to as probing the spin structure of the proton, as for free fields the axial charge is related to spin.

So far the discussion has been generic. We will now restrict ourselves to a particularly interesting limit in which DIS is often considered, the so-called Bjorken limit. In this case, the momentum transfer $Q^2$ is sent to infinity while keeping $x_B$ fixed. Asymptotic freedom then suggests that the physics can be described in terms of free quarks and gluons. This results in the structure functions depending only on $x_B$, a phenomenon known as ``Bjorken scaling". This is indeed the case at leading order, and the OPE and other perturbative QCD methods can be used to systematically compute corrections. Scaling, taken literally,  leads to the ``naive parton model", where DIS cross sections are computed by considering a lepton scattering off a quark which carries a momentum fraction $\xi$ of the proton.

The parton model is usually formulated in a frame where the target nucleon has infinite momentum, the so-called infinite momentum frame (IMF). In this frame the target mass can be neglected, so that the nucleon has 4-momentum (assuming that it moves in the $z$-direction) $p^\mu = (p,0,0,p)$ and the parton has exactly a fraction $\xi$ of this momentum: $(\xi p,0,0,\xi p)$. By computing the scattering of a virtual photon and a free quark, represented in figure  \ref{fig:DISParton} (see for instance \cite{Manohar:1992tz}), one can easily find the contribution of a single quark with momentum fraction $\xi$ to the distribution functions

\begin{align}
  F^{q,\xi}_1(x_B) &= \frac{e_q^2}{2}\delta(\xi - x_B) \\
  F^{q,\xi}_2(x_B) &= x_B e_q^2\delta(\xi - x_B) = 2x_B   F^{q,\xi}_1(x_B) \\
  g^{q,\xi}_1(x_B) &= \frac{e_q^2}{2} h_q h_H \delta(\xi - x_B) \\
  g^{q,\xi}_2(x_B) &= 0 \ ,
\end{align}
where $e_q$ is the electric charge of the quark and $h_q$ is the helicity of the free quark. The quantity $h_H$ denotes the helicity of the target; in the IMF polarization of the target becomes $s_H=ph_H$. In this picture, the structure functions acquire an intuitive meaning. Defining $q_\pm(\xi)$ as being the probability distribution of a quark whose helicity is aligned/antialigned to the proton's,  integrating over the momentum fraction $\xi$ and summing over helicities, we obtain
\begin{align}
  F^{part.\ mod.}_1(x_B) &= \sum_q \frac{e_q^2}{2}\left(q_+(x_B) + q_-(x_B) +\bar q_+(x_B) +\bar q_-(x_B)\right) \\
  F^{part.\ mod.}_2(x_B) &=  2 x F_1(x_B)  \\
  g^{part.\ mod.}_1(x_B) &= \sum_q \frac{e_q^2}{2}\left(q_+(x_B) - q_-(x_B) +\bar q_+(x_B) -\bar q_-(x_B)\right) \\
  g^{part.\ mod.}_2(x_B) &= 0 \ .
\end{align}

We see that in the Bjorken limit, $x_B$ can be interpreted as the momentum fraction of a proton that a constituent parton is carrying in the infinite momentum frame. The function $F_1(x_B)$ is then simply the probability distribution of finding a given quark in the hadron with momentum fraction $x_B$. The function $g_1(x_B)$ measures polarization asymmetries in the quark distributions.

The integral of $g_1$ over $x_B$ yields its first Mellin moment;  plugging in the quark charges, we get
\begin{align}
  \int \dd x_B g^{part.\ mod.}_1(x_B) &= \frac{1}{2}\left[\frac{4}{9} \Delta u + \frac{1}{9} \Delta d + \frac{1}{9} \Delta s \right] ,
\end{align}
where we have defined $\Delta q = \int \dd x_B\left( q_+(x_B) - q_-(x_B)\right)$. It is useful to further rewrite it in terms of axial currents that appear in  eq.~(\ref{eq:OPEg1}). The axial charges are defined as
\beq
2 M s^\mu g_A^{(a)} = \langle \proton,s| j_5^{\mu (a)}| \proton,s\rangle.
\eeq
A frequently used convention is  to reabsorb some of the $SU(3)$ normalization into the axial charges by defining \cite{Shore:2007yn}
\begin{align}
a^3 = 2 g_A^3, \ \ \ \  a^8 = 2\sqrt{3} g_A^8, \ \ \ \ a^0 = \sqrt{6} g_A^0 \ .
\end{align}
In these terms, the parton model prediction for the first moment of $g_1$ yields the Ellis-Jaffe sum rule
\begin{align}
  \int_0^1 \dd x_Bg_1(x_B)  =  \frac{1}{12} a^{3} + \frac{1}{36} a^{8} + \frac{1}{9} a^{0} \ ,
\end{align}
 where the axial charges $a^3$ and $a^8$ can be independently inferred from low-energy data on neutron $\beta$-decay and hyperon decay (assuming $SU(3)$ symmetry) \cite{Ellis:1973kp}. Assuming the validity of the parton model, a measurement of the first moment of $g_1$ is effectively a measurement of the singlet axial charge $a^0 = \Delta u + \Delta d + \Delta s$.

 The measurements of polarization asymmetries in the final states in polarized DIS give a value of $a_0|_{Q^2\to\infty} = 0.33 \pm 0.06 $ \cite{Shore:2007yn}. This value is in sharp contradiction with the value predicted by the constituent quark model. As reviewed in   \cite{Bass:2004xa}, the most na\"ive nonrelativistic $SU(6)$ description of the proton yields $a^0=1$. More realistic relativistic ``bag" models predict $a^0 \approx 0.6$, still leaving a discrepancy of a factor of 2.

A major ingredient which has been missing in our discussion so far is how the gluons contribute to the proton axial charge. In the partonic description, the gluons are absent from the picture. This can be traced back to the OPE for $g_1$ \eqref{eq:OPEg1}, where no gluonic operators are present at leading order. Indeed, the only gluonic operator of twist two is the topological charge density
\begin{align}
  Q(x) = \frac{\alpha_s}{4\pi} \Tr\left(G_{\mu\nu}\tilde G^{\mu\nu}\right)
\end{align}
where $\alpha_s$ is the QCD coupling constant. As is well known, this operator is a total derivative
\begin{align}
  Q(x) &= 2  \dr_\mu K^\mu\\
  K_\mu &= \frac{\alpha_s}{8\pi}\epsilon_{\mu\nu\lambda\rho} \left[A^\nu_a\left(\partial^\lambda A^\rho_a - \frac{1}{3}g f_{abc} A^\lambda_b A^\rho_c\right)\right] \ ,
\end{align}
and as a result does not contribute in perturbation theory. It is also well known that topological charge is precisely the operator that nonperturbatively leads to an anomalous nonconservation of the axial vector
\begin{align}
\partial_\mu j_5^\mu = N_f Q(x) \label{eq:axialanomalyQCD}
\end{align}
where $N_f=3$ is the number of flavors. As a result the operator $Q(x)$ is expected to give a substantial nonperturbative contribution to the axial charge. To describe this effect, we can rewrite the anomaly equation \eqref{eq:axialanomalyQCD} as
\beq
\dr_\mu(j_5^\mu - 2 N_f K^\mu) = 0 \ .
\eeq
This form suggests to introduce a similar decomposition for the axial charge. In particular, we decompose $a^0$ as
\begin{align}
  a^0 = \tilde a^0 - \frac{\alpha_s}{2\pi} N_f \Delta g \ ,
\end{align}
with $\Delta g$ an explicit contribution from the topological gluonic operator and $\tilde a^0$ the remainder of the  proton axial charge.

This decomposition was suggested early on as a solution to the ``spin" crisis \cite{Carlitz:1988ab,Altarelli:1988nr,Hatsuda:1988jv,Jaffe:1989jz,Shore:1990zu}, and recently revisited in   \cite{Tarasov:2020cwl,Tarasov:2021yll}. Indeed, as suggested by this decomposition, the topological gluonic configuration can potentially screen the axial charge inside the proton. In this scenario, one expects that $\tilde a_0$ corresponds to the net axial charge carried by quarks inside the proton, in accord with the na\"ive quark model.


%

\end{document}